\documentclass[a4paper,11pt]{article}
\RequirePackage{latexsym,amsmath,amssymb}
\usepackage{graphicx,color}

\usepackage{fullpage}
\usepackage[british]{babel}

\usepackage[T1]{fontenc}
\usepackage{hyperref}
\newcommand{\lp}{\ell_{\rm p}}
\newcommand{\mpl}{m_{\rm p}}
\newcommand{\gn}{G_{\rm N}}
\newcommand{\Nb}{N_{\rm B}}
\newcommand{\ab}{a_{\rm B}}
\newcommand{\Rb}{R_{\rm B}}
\newcommand{\mb}{m_{\rm B}}
\newcommand{\DE}{{\rm DE}}
\newcommand{\DF}{{\rm DF}}
\newcommand{\ppar}{{p_\parallel}}
\newcommand{\port}{{p_\perp}}
\newcommand{\Neff}{N_{\text{eff}}}
\newcommand{\rmd}{\mathrm{d}}
\newcommand{\e}[1]{\operatorname{e}^{#1}}
\newcommand{\be}{\begin{eqnarray}}
\newcommand{\ee}{\end{eqnarray}}
\numberwithin{equation}{section}
\begin{document}
\begin{center}
{\Large \textbf{Effective Fluid Description of the Dark Universe}}\\[2em]
\renewcommand{\thefootnote}{\fnsymbol{footnote}}
M.~Cadoni${}^{a,b}$\footnote[1]{E-mail: mariano.cadoni@ca.infn.it},
R.~Casadio${}^{c,d}$\footnote[2]{E-mail: casadio@bo.infn.it},
A.~Giusti${}^{c,d,e}$\footnote[3]{E-mail: agiusti@bo.infn.it},
W.~M{\"u}ck${}^{f,g}$\footnote[4]{E-mail: mueck@na.infn.it},
and
M.~Tuveri${}^{a,b}$\footnote[5]{E-mail: matteo.tuveri@ca.infn.it}
\renewcommand{\thefootnote}{\arabic{footnote}}
\\[1em]
{\footnotesize \em
${}^a$Dipartimento di Fisica, Universit\`a di Cagliari, Cittadella Universitaria, 09042 Monserrato, Italy\\
${}^b$I.N.F.N, Sezione di Cagliari, Cittadella Universitaria, 09042 Monserrato, Italy\\
${}^c$Dipartimento di Fisica e Astronomia, Universit\`a di Bologna, via Irnerio~46, 40126 Bologna, Italy\\
${}^d$I.N.F.N., Sezione di Bologna, IS - FLAG, via B.~Pichat~6/2, 40127 Bologna, Italy\\
${}^e$Arnold Sommerfeld Center, Ludwig-Maximilians-Universit\"at, Theresienstra\ss e~37,~80333 M\"unchen, Germany\\
${}^f$Dipartimento di Fisica ``Ettore Pancini'', Universit\`a di Napoli ``Federico II'', via Cintia, 80126 Napoli, Italy\\
${}^g$I.N.F.N., Sezione di Napoli, via Cintia, 80126 Napoli, Italy\\[2em]
}
\begin{abstract}
We  propose an effective anisotropic fluid description for a generic 
infrared-modified theory of gravity.
In our framework, the additional component of the acceleration, commonly 
attributed to dark matter, is explained as a radial pressure generated by 
the reaction of the dark energy fluid to the presence of baryonic matter.
Using quite general assumptions, and a microscopic description
of the fluid in terms of a Bose-Einstein condensate of gravitons, we 
find the static, spherically symmetric solution for the metric in terms 
of the Misner-Sharp mass function and the fluid pressure. At galactic 
scales, we correctly reproduce the leading MOND-like $\log(r)$ and 
subleading $(1/r)\,\log( r)$ terms in the  weak-field expansion of the 
potential. Our description also predicts a tiny (of order $10^{-6}$ for 
a typical spiral galaxy) Machian modification of the Newtonian potential 
at galactic scales, which is controlled by the cosmological acceleration. 
\end{abstract}
\end{center}
\section{Introduction}
\label{s:intro}
One of the most intriguing puzzles of contemporary fundamental physics
is the origin of the dark components of matter and energy in our
universe~\cite{Zwicky:1933gu, Faber:1979pp, deSwart:2017heh, Peebles:2002gy}. 
The most conservative approach to this problem, the $\Lambda$CDM 
model~\cite{Penzias:1965wn, Ade:2013zuv}, explains the experimental data 
about the present accelerated expansion of the universe~\cite{Riess:1998cb},
the structure formation, the galaxy rotation curves and gravitational lensing
effects~\cite{Rubin:1980zd, Persic:1995ru, Massey:2010hh},
by assuming that about $95\%$ of the matter of our universe is exotic.
\par
Despite the extensive agreement with large scale structure and cosmic 
microwave background observations, 
the $\Lambda$CDM model is not completely satisfactory, not only from 
a conceptual point of view, but also because there is some tension at the level
of the phenomenology of galaxies and galaxy clusters.  
Concerning the Milky way galaxy, for example, three problems arise: 
the {\it missing satellite problem}~\cite{Klypin:1999uc,Moore:1999nt}
($N-$body simulations predict too many dwarf galaxies within the Milky 
Way virial radius), the {\it cusp-core problem}~\cite{DeBlock:2010} 
(too much dark matter in the innermost regions of galaxies w.r.t.~observations) and the 
{\it too-big-to-fail problem}~\cite{BoylanKolchin:2011de,BoylanKolchin:2011dk} 
(the dynamical properties of the most massive satellites in the 
Milky way are not correctly predicted by simulations).
In particular, these problems become more and more evident when 
one tries to study galaxy rotation curves.
Typically, the rotational velocity in galaxies approaches a non-zero asymptotic value
with increasing distance from a galaxy's centre.
This asymptotic value satisfies an empirical relationship with the galaxy's total luminosity
known as the Tully-Fisher relation~\cite{Tully:1977fu}.
Rephrased as a relation between the asymptotic velocity $v$ and the 
total baryonic mass $\mb$, it takes the form $\mb \sim v^4$ (baryonic 
Tully-Fisher relation)~\cite{McGaugh:2000sr, McGaugh:2012}.
With adjusted units, it is equivalent to~\footnote{We use units with $c=1$, while the Newton 
and Planck constants are expressed in terms of the Planck length and mass as
$\gn=\lp/\mpl$ and $\hbar= \lp\mpl$, respectively.}
\begin{equation}
\label{tully-fisher}
	v^2
	\approx
	\sqrt{a_0\, \gn\, \mb}
	\ ,
\end{equation} 
where $a_0$ denotes an empirically determined factor with dimensions of an acceleration.
The surprising fact is that the value of $a_0$ appears to be $a_0\approx H/(2\,\pi) \approx H/6$,
where  $H$ is the current value of the Hubble constant.
This coincidence begs for a deeper physical explanation and points to a deep 
connection between the dark matter and dark energy (DE) phenomena.
\par
To explain the Tully-Fisher relation within a $\Lambda$CDM model, one must 
assume that the dark matter halos of all galaxies contain just the right 
amount of dark matter, which is obviously not a physically motivated 
assumption.
For this reason, the Tully-Fisher relation has been used to argue in support of modified
theories of gravity, where the standard description of the gravitational interaction given 
by Einstein's general relativity~(GR) is modified at large scales.
The departure from GR in such alternative approaches may involve 
modifications of the Einstein-Hilbert action, like in $f(R)$ 
theories~\cite{Capozziello:2011et,Nojiri:2017ncd}, string inspired 
brane-world scenarios~\cite{Dvali:2000hr}, or a change of 
the paradigm that describes gravity by means of a metric and 
covariant theory. To this last class of approaches belongs Milgrom's 
Modified Newtonian dynamics~(MOND)~\cite{Milgrom:1983ca,Milgrom:2014usa}. 
In the MOND framework, in which the acceleration $a_0$ is promoted to 
a fundamental constant, the gravitational acceleration is modified 
with respect to its Newtonian form. At distances outside a 
galaxy's inner core, it reads 
\begin{equation}
\label{MOND}
	a_{\rm MOND}(r)
	=
	\sqrt{a_0\, a_{\rm B}(r)}
	\ ,
\end{equation}
where $a_{\rm B}(r) = \gn \mb(r)/r^2$ is the Newtonian radial acceleration
that would be caused by the baryonic mass $\mb(r)$ inside the radius $r$.
Phenomenologically, the simple formula~\eqref{MOND} turns out to explain 
the rotation curves of galaxies surprisingly well~\cite{McGaugh:2008nc, McGaugh:2012}, 
although it cannot explain the mass deficit in galaxy clusters~\cite{Sanders:1998kv}.
More recently, Verlinde~\cite{Verlinde:2016toy} has given a
controversial~\cite{Milgrom:2016huh, Dai:2017qkz} derivation of the MOND
formula~\eqref{MOND}, proposing that the dark matter phenomena can be attributed 
to an elastic response of the DE medium permeating the universe.
\par
One common problem of these approaches is the difficulty of performing a 
``metric-covariant uplifting'' of the theory~\cite{Hossenfelder:2017eoh,Dai:2017qkz}.
In fact, such theories are usually formulated in the weak-field regime, 
whereas we know that gravity must allow for the metric-covariant description 
given by GR, at least at solar system scales.
Fluid space-time models may provide a simple way to perform such an 
uplifting.
For example, it is well known that de~Sitter space is 
equivalent to the space-time of an isotropic fluid with constant energy 
density and equation of state $p=-\varepsilon$. Phenomenologically, 
galaxy rotation curves and gravitational lensing have been described 
using two-fluid~\cite{Harko:2011kw, Harko:2011nu} and anisotropic 
fluid models~\cite{Bharadwaj:2003iw, Faber:2005xc, Su:2009fd}.
It is also possible to extend such models to contain DE~\cite{Zhao:2008vua},
although the physical nature of these fluid models has yet to be established.
\par
In this letter, we propose a way to describe the infrared modification of gravity 
remaining in a GR framework, by codifying it in terms of an effective 
(anisotropic) fluid acting as a source in the Einstein equations.
We focus on the dark matter phenomenology within a single galaxy in a 
background de~Sitter space-time.
We will not address the problem of explaining the origin of DE, the existence
of which we take for granted and which we describe by means of a DE fluid
component with vacuum equation of state
$\varepsilon_{\DE} = - p_{\DE}={3\,H^2}/({8\,\pi\, \gn})$.
For small velocities, such a system is approximately described 
by a static, spherically symmetric geometry, whose physical content 
is effectively represented by an anisotropic fluid.
The spherical symmetry should be considered as a first rough approximation 
for spiral and elliptical galaxies, which we adopt in the light 
of the fact that more realistic anisotropic fluid solutions 
with rotational symmetry are not known explicitly at the moment.
\par
This model easily accommodates the observed deviation of the 
galaxy rotation curves from the Newtonian prediction at a typical 
infrared scale $r_0\sim\sqrt{\gn\, \mb/H}$.
These deviations are commonly attributed to dark matter, but in fact,
they only imply the existence of some \emph{dark force}.
Taking a viewpoint similar to Verlinde's~\cite{Verlinde:2016toy},
we propose that this dark force can be entirely ascribed to the
back-reaction of the DE fluid to the presence of baryonic matter
and, therefore, is completely determined by the distribution of the
latter.
In contrast to Verlinde's, however, we explore the possibility that 
this back-reaction leads to an effective pressure term and not 
an effective mass density in the anisotropic fluid description. 
Since this pressure term is, {\em a priori\/}, an arbitrary function 
of the radial distance from the galactic centre, we need some input
from an underlying microscopic theory in order to make our model
predictive.
We will employ a microscopic description in terms of a corpuscular picture,
in which the DE fluid component is given by a Bose-Einstein condensate 
of gravitons~\cite{Dvali:2011aa, Dvali:2013eja}, whereas the 
back-reaction effects are carried by gravitons in the non-condensed 
phase of the fluid. This will give us a way to relate the pressure 
causing the dark force to the Newtonian acceleration originating 
from the presence of baryonic matter.
\par
Finally, we will calculate the effective metric components for our model 
and show that they contain, in the weak-field approximation, 
the typical $\log (r/r_0)$ MOND gravitational potential at 
galactic scales. Moreover, at such scales, we will also find a 
tiny Machian correction to the Newtonian potential depending 
on the position of the cosmological horizon.  
\section{Anisotropic fluid space-time}
\label{s:fluid}
We start by considering a static, spherically symmetric system, 
for which one can employ the Schwarzschild-like metric
\begin{equation}
\label{fluid:metric}
	\rmd s^2 = -f(r) \e{\gamma(r)} \rmd t^2 + \frac{\rmd r^2}{f(r)} +r^2 \rmd \Omega^2
	\ .
\end{equation}
It is known that this metric is, in all generality, a solution 
to Einstein's equations with the energy-momentum tensor of an 
anisotropic fluid~\cite{Cosenza1981, Herrera1997},
\begin{equation}
\label{fluid:em.tensor}
T^{\mu\nu}
=
\left(\varepsilon+\port\right)
u^\mu u^\nu
+\port g^{\mu\nu}
-\left(\port -\ppar\right)
v^\mu v^\nu
~,
\end{equation}
where the vectors $u^\mu$ and $v^\mu$ satisfy $u^\mu\, u_\mu = -1$,
$v^\mu\, v_\mu = 1$, and $u^\mu\, v_\mu =0$.
Explicitly, the fluid velocity is $u^\mu = \left( f^{-1/2}\, \e{-\gamma/2},0,0,0\right)$ and 
$v^\mu = \left( 0,f^{1/2} ,0,0\right)$ points radially outwards.
The energy density is given by $\varepsilon$, and $\port$ and 
$\ppar$ denote the pressures perpendicular and parallel to the 
space-like vector $v^\mu$, respectively.
Energy-momentum conservation is equivalent to the hydrostatic equilibrium condition,
and imposes constraints on these quantities. 
\par
The Einstein equations with the energy-momentum tensor~\eqref{fluid:em.tensor} 
are solved by
\begin{subequations}
\begin{align}
f(r)
&=
1-\frac{2\,\gn\,m(r)}{r}
\ ,
\label{fluid:f.sol1}
\\
\gamma'(r)
&=
\frac{8\,\pi\,\gn\, r}{f(r)}
\left(\varepsilon+\ppar \right)
\ ,
\label{fluid:f.sol2}
\end{align}
where primes denote differentiation with respect to~$r$, and
\begin{equation}
m(r)
=
4\,\pi
\int_0^r 
\rmd \tilde{r}\, {\tilde{r}}^2\, \varepsilon(\tilde{r})
\label{fluid:f.sol3}
\end{equation}
\end{subequations}
is the Misner-Sharp mass function representing the total energy 
inside a sphere of radius $r$. Finally, the tangential pressure 
follows from energy-momentum conservation,
\begin{equation}
\label{fluid:em.pperp}
\port
=
\ppar
+\frac{r}2
\left[ \ppar'
+ \frac{1}{2} \left(\varepsilon +\ppar \right)
\left( \frac{f'}f +\gamma' \right)
\right]
\ .
\end{equation}
\par
Let us then consider a test particle comoving with the fluid, so that its 
four-velocity is $u^\mu$.
The four-acceleration necessary to keep it at a fixed coordinate radius $r$ is given
by $a^\mu=u^\nu\, \nabla_\nu u^\mu$.
In the frame of Eq.~\eqref{fluid:metric}, only the radial component of this
acceleration does not vanish and is given by
\begin{equation}
\label{fluid:ar.final}
a^r
\equiv
a
=
\frac12 \left( f\,\gamma' + f' \right)
=
\frac{\gn\,m(r)}{r^2}
+
4\, \pi\, \gn\, r\, \ppar(r)
\ . 
\end{equation}
In Newtonian language, the first term has the obvious interpretation 
as the acceleration that counters the gravitational pull of the central mass. 
The second term may be interpreted as the acceleration caused by the radial
pressure. 
The same result can be obtained by considering the geodesic motion along
a circular orbit of radius $r$, with $\theta=\pi/2$ and constant angular velocity
$\Omega= \rmd \phi/\rmd t$. 
Of course, this is the physically relevant situation for the motion of 
stars within a galaxy.
Starting with the four-velocity $u^\mu = C(r) \left( 1,0,0,\Omega\right)$, with $C(r)$
such that $u_\mu u^\mu =-1$, and solving the geodesic equation at fixed $r$ and 
$\theta=\pi/2$, one obtains $\Omega^2 = \e{\gamma}{a}/r$,
with $a$ again given by Eq.~\eqref{fluid:ar.final}.
\par
The above equations can describe a variety of physical situations. 
De~Sitter space is equivalent to an isotropic DE fluid with the constant energy density
$\varepsilon_{\DE}$ and pressure $\ppar_{\DE}=\port_{\DE}=p_{\DE}$ satisfying
\be
\varepsilon_{\DE} = - p_{\DE}
=
\frac{3\,H^2}{8\,\pi\, \gn}
\ .
\label{eps.p.dS}
\ee
This yields 
\be
f(r) = 1-H^2\,r^2
\ ,
\label{dS}
\ee
with $\gamma=0$, and 
\begin{equation}
\label{ar.DS}
	a_{\DE}(r) = - H^2\, r
	\ .
\end{equation}
Being maximally symmetric, de~Sitter space does not 
allow for circular geodesics, which is confirmed by the fact that 
$a_{\DE}$ is negative. 
This acceleration describes the accelerating cosmological expansion
of the universe.
Notice that, because of its vacuum equation of state~\eqref{eps.p.dS}, the 
DE fluid component does not contribute to $\gamma$ but enters only in $f$ via 
the de~Sitter term.
\par
Pressureless baryonic matter can be easily added to de~Sitter space, 
\begin{equation}
\label{eps}
	\varepsilon = \varepsilon_{\rm B} +\varepsilon_{\DE}
	\ ,
\end{equation}
where $\varepsilon_{\DE}$ is again given in Eq.~\eqref{eps.p.dS}.
The Misner-Sharp mass function will split correspondingly, 
\be\label{eps1}
m(r)
=
\mb(r)+m_{\DE}(r)
=
\mb(r)+\frac{H^2\, r^3}{2\, \gn}
\ .
\label{msplit}
\ee
and the metric function $f$  turns out to be
\be
f(r)= 1-\frac{2\,\gn\, \mb(r)}{r} -H^2\, r^2.
\label{SdS}
\ee
This leads to a Newtonian acceleration term
\begin{equation}
\label{ar.B}
	\ab(r) = \frac{\gn\,\mb(r)}{r^2}
	\ ,
\end{equation}
in addition to \eqref{ar.DS}.
If the baryonic matter is localized within a radius $\Rb$ then, for $r>\Rb$, the space-time 
is identical to the Schwarzschild-de~Sitter solution.  
\par
The observed galaxy rotation curves imply that, in addition to $a_\DE$ 
(which, in this context, is actually negligible) and $\ab$, there is an acceleration
caused by a dark force,
\begin{equation}
\label{a.DF.intro}
	a = \ab + a_\DE + a_\DF
	\ .
\end{equation}
We think that dark matter does not exist as an independent form of matter,
but rather that the phenomena usually attributed to it are a consequence of
the interaction between the baryonic matter and the DE fluid.
We therefore assume the energy density and the Misner-Sharp in the 
cosmos are given respectively  by Eqs.~(\ref{eps}) and~(\ref{eps1}).
Taking the baryonic matter as approximately pressureless, we write 
\begin{equation}
\label{ppar}
	\ppar = \ppar_{\DE} + \ppar_{\DF}
	\ ,
\end{equation}
where $\ppar_\DF$ is the pressure that generates the dark force.
In the next section, we will derive $\ppar_{\DF}$ from the point of view of
a corpuscular interpretation of gravity in general, and of the de~Sitter space
in particular. 
\par
At galactic scales, we can neglect the DE terms $p_{\DE}$ and $\varepsilon_{\DE}$.
Splitting 
the total radial gravitational acceleration into the baryonic acceleration 
$\ab$ and the dark acceleration $a_{\DF}$,
Eq.~\eqref{fluid:ar.final} now gives
\begin{equation}
\label{poi}
\ab + a_{\DF}
\simeq
\frac{\gn\,\mb(r)}{r^2}
+4 \,\pi\, \gn\, r\, \ppar_{\DF}(r)
 \ .
\end{equation}
The first term on the right hand side is exactly $\ab$, thus 
the dark acceleration is completely due to the pressure of the 
anisotropic fluid.
This is an important point, because it implies that the modifications to GR
at galactic scales commonly attributed to dark matter
can be generated by the pressure $\ppar$ in our effective fluid description.
Since this pressure term can be thought of as a reaction of the DE fluid to
the presence of baryonic matter, it is conceptually very similar to Verlinde's
description of dark forces as the elastic response of the DE medium to the
presence of baryonic sources~\cite{Verlinde:2016toy}.
Note also that $\ppar_{\DF}$ will necessarily give rise to an anisotropic
component $\port_{\DF}$ according to the conservation Eq.~\eqref{fluid:em.pperp}.
\section{Corpuscular dark force}
\label{ss:cdf}
In this section, we review the fundamentals of the  corpuscular picture of the
de~Sitter space, express the accelerations $a_\DE$ (DE without matter)
and $\ab$ (Newtonian acceleration) in terms of corpuscular quantities
and derive $a_\DF$ from the corpuscular picture of de~Sitter in the
presence of baryonic matter. 
We anticipate that the result will be the MOND formula~\eqref{MOND} 
(up to a multiplicative constant).
Throughout this section, numerical factors of order unity will mostly be omitted.
\par
The basis of the corpuscular picture of gravity~\cite{Dvali:2011aa, Dvali:2013eja} 
is that the classical gravitational field of an (isolated) object of mass $m$ is in
fact a quantum coherent state of gravitons with occupation
number~\cite{Mueck:2013mha,Casadio:2016zpl,Casadio:2017cdv} 
\begin{equation}
\label{N.def}
	N \sim \frac{m^2}{\mpl^2}
	\ .
\end{equation}
These gravitons are closely bound to the source and interact with 
other objects nearby, \emph{e.g.}, a test particle.
If $r$ is the distance between the test particle and the massive object, the effective 
interaction energy for each graviton is $\omega(r)=\hbar/r$. 
Therefore, we can express the Newtonian gravitational acceleration felt
by the test particle in terms of $\omega(r)$ and $N$ as
\begin{equation}
\label{aNewt}
	a(r) = \frac{\gn\, m}{r^2}
	\sim
	\frac{\omega^2(r)}{\mpl^2\,\lp}\, \sqrt{N}
	\ . 
\end{equation}
The argument generalises straightforwardly to a spherically symmetric distribution
of mass.
In this case, however, not all gravitons can contribute to the acceleration of the test particle,
but only those that are bound to the mass inside the radius $r$. Henceforth, let us denote by $\Neff(r)$ the effective number of gravitons which contribute to the acceleration of a test particle at radius $r$.\footnote{The number $\Neff(r)$ is not a 
good classical observable and must not be confused with the number of gravitons inside the radius $r$. Such a number does not exist, because, relativistically, there is no notion of a local number density.} 
In the case at hand, it is $\Neff(r)=m^2(r)/\mpl^2$, and \eqref{aNewt} becomes
\begin{equation}
\label{aNewt2}
	a(r) = \frac{\gn\, m(r)}{r^2}
	\sim
	\frac{\omega^2(r)}{\mpl^2\,\lp} \sqrt{\Neff(r)}
	\ . 
\end{equation}
In the above argument it is important that the gravitons are in the normal
(non-condensed) phase, for which we can use the effective law
$\omega(r)=\hbar/r$. 
We shall call {\em corpuscular acceleration\/} the quantity
\begin{equation}
\label{a.corpus}
	a(r) \sim \frac{\omega^2(r)}{\mpl^2\,\lp} \sqrt{\Neff(r)}
	\ .
\end{equation}
Although we have derived this formula for non-condensed gravitons, 
which generate the Newtonian acceleration, it turns out to be valid also for the acceleration
caused by condensed gravitons, as we will verify in the following.
\par
The DE fluid of the pure de~Sitter space-time~\eqref{dS} is described in the corpuscular
picture~\cite{Dvali:2013eja,Casadio:2015xva,Casadio:2017twg} as
a Bose-Einstein condensate of $N$ (very soft and virtual) gravitons with typical
energy $\omega=\hbar\,H$.
Since the total energy of the DE fluid inside the de~Sitter horizon of radius $1/H$
is given by $m_H=1/({2\,\gn\,H})$, one has~\footnote{Similar relations hold for the
case of a Schwarzschild black hole~\cite{Dvali:2011aa}.}
\begin{equation} 
\label{scaling}
	N
	\sim
	\frac{m^2_H}{\mpl^2}
	\sim
	\frac{1}{\lp^2\,H^2}
	\ ,
\end{equation} 
with $N\,\omega = m_H$.
It is important that the number of gravitons scales holographically with the horizon size.
Consider a test particle at a fixed distance $r$.
As we recalled in the previous section, such a particle is not in geodesic motion,
but feels the acceleration~\eqref{ar.DS} caused by the DE condensate.
Let us check whether the corpuscular acceleration formula~\eqref{a.corpus}
reproduces this result.
In order to estimate $\Neff(r)$, \emph{i.e.}, the effective number of gravitons in the condensate that 
contribute to the interaction with the test particle, we use the fact that the graviton number
scales holographically (with area) and all gravitons contribute to the acceleration of a test particle, 
when it is at the horizon.
In other words, $\Neff(r)$ must match Eq.~\eqref{scaling} for $r=1/H$.    
This leads to
\begin{equation} 
\label{Nr}
	\Neff(r) \sim \frac{r^2}{\lp^2}
	\ .
\end{equation} 
Moreover, since the gravitons are now in the condensed phase,
the interaction energy $\omega(r)=\omega=\hbar\,H$ is constant. 
Therefore, Eq.~\eqref{a.corpus} yields
\begin{equation}
\label{aDE.N}
	|a_\DE(r)|
	\sim \frac{\omega^2}{\mpl^2\,\lp} \sqrt{\Neff(r)}
	=
	H^2\, r
	\ ,
\end{equation}
which reproduces the expected result~\eqref{ar.DS}. 
\par
Putting the above arguments together leads to a new effect. 
Let us consider baryonic matter present in a relatively small amount
(say $\mb\ll m_H$) and localised within some radius $\Rb$. 
The space-time will be given by the Schwarzschild-de Sitter 
solution with $f(r)$ in Eq.~\eqref{SdS} for $r >\Rb$.
Note, in particular, that the horizon radius $L$ is now determined
by the corresponding $f(L)=0$, \emph{i.e.},
\begin{equation}
\label{SdS.horizon}
	H^2\, L^2
	=
	1 -\frac{2\,\gn\,\mb}{L}
	=
	1-\frac{\mb}{m(L)}
	\ ,
\end{equation}  
where $m(L)$ denotes the total (Misner-Sharp) mass of the space-time,
\be
m(L)
=
\frac{L}{2\,\gn}
\ .
\label{eq:mL}
\ee
From the corpuscular point of view, the DE fluid will react to 
the presence of baryonic matter, but, since $\mb\ll m(L)$,
most of the gravitons will remain in the condensed phase and retain an energy 
$\omega\sim \hbar/L$. 
From Eqs.~\eqref{N.def} and \eqref{SdS.horizon}, their number is given by 
\begin{equation}
\label{N0}
	N_\DE
	\sim
	\frac{\left[m(L)-\mb\right]^2}{\mpl^2}
	\sim
	\frac{H^4\, L^4\,m^2(L)}{\mpl^2}
	\sim
	\frac{H^4\, L^6}{\lp^2}
	\ .
\end{equation}
Using the same reasoning as above, the number of the condensed gravitons that
effectively contribute to the cosmological  acceleration of a test particle at radius $r$ is $\Neff{}_{,\DE}(r)=H^4 \,L^4\, r^2/\lp^2$,
and Eq.~\eqref{aDE.N} remains valid.
\par
However, according to Eq.~\eqref{N.def}, the total number of gravitons in the system is given
by 
\be
\label{Ntot}
N
\sim
\frac{m^2(L)}{\mpl^2}
\ .
\ee
This implies that there are $N-N_\DE$ gravitons, which are not in the condensed phase and, therefore,
behave differently from the condensate.
Since, from Eqs.~\eqref{N0} and \eqref{Ntot},
\begin{equation}
\label{N1}
	N-N_\DE
	\sim
	\frac{L\,\mb}{\lp\, \mpl} - \frac{\mb^2}{\mpl^2}
	\ ,
\end{equation}
there must be many more non-condensed gravitons than those that are closely bound to the baryonic mass.
In fact, the number of the latter is simply
\begin{equation}
\label{NB}
	\Nb \sim \frac{\mb^2}{\mpl^2}
	\ ,
\end{equation} 
and their contribution to the acceleration is the Newtonian term~\eqref{ar.B}.
The remaining non-condensed gravitons, with total number~\footnote{The hierarchy
of the graviton numbers is $\Nb\ll N_\DF\ll N_\DE$.
There must also be corrections, sub-leading in $\gn \mb/L$, to account for the second
term with the negative sign in \eqref{N1}.}
\begin{equation}
\label{NDF}
	N_\DF
	\sim
	\frac{L\, \mb}{\hbar}
	\ ,
\end{equation}   
mediate the interaction between the baryonic matter and the DE 
condensate.
\par
In order to estimate the effective number of these non-condensed gravitons that contribute to the
acceleration of a test particle at the radius $r$, $\Neff{}_{,\DF}(r)$, we note that the overall
scaling is again holographic, but we must also take into account that
only those gravitons that are ``pulled out'' of the condensate by the baryonic mass
inside the radius $r$ can contribute (if $\mb$ were constant, it would be simply
holographic).
Hence,
\begin{equation}
\label{NDF.r}
	\Neff{}_{,\DF}(r)
	\sim
	\frac{r^2\, \mb(r)}{\hbar\, L}
	\ .
\end{equation}   
Finally, from Eq.~\eqref{a.corpus} with $\omega(r)=\hbar/r$ (for non-condensed gravitons),
we obtain 
\begin{equation}
\label{aDF.N}
	|a_\DF(r)| \sim \sqrt{\frac{\gn\,\mb(r)}{L\,r^2}}
	\sim
	\sqrt{\frac{\ab(r)}{L}}
	\ , 
\end{equation}
which is precisely the MOND acceleration~\eqref{MOND} up to a numerical 
factor.
Therefore, the corpuscular picture naturally explains the 
presence of a dark force and the approximate coincidence of the 
MOND acceleration $a_0$ with the Hubble constant $H \approx 1/L$. 
This is our main result in this section. 
Moreover, the pressure necessary to sustain the dark force is given by
\begin{equation}
\label{ppar.DF}
 	\ppar_\DF
	\sim
	\frac{1}{4\,\pi\, r^2}\, \sqrt{\frac{\mb(r)}{\gn\,L}}
	\ ,
\end{equation}
as follows from Eqs.~\eqref{poi} and \eqref{aDF.N}.
\par
Let us conclude this section with a few remarks.
First, the previous arguments give order-of-magnitude estimates only, 
without precise numerical factors and without information on 
the directions of the various contributions to the acceleration. 
Second, all expressions must receive higher order corrections 
in $ \gn\,\mb/L$, as can be seen, \emph{e.g.}, from the different 
signs of the two terms in Eq.~\eqref{N1}.
Presumably, these corrections will be responsible for the cross-over between
the Newtonian and  the MOND regimes as well as between the MOND and
the de Sitter regimes. 
\section{Metric at galactic scales}
\label{ss:galactic}
Starting from Eqs.~\eqref{eps}, \eqref{ppar} and \eqref{ppar.DF}, we will 
now evaluate the metric of the anisotropic fluid space-time. For any 
given distribution of baryonic matter $\varepsilon_{\rm B}=\varepsilon_{\rm B}(r)$,
Eqs.~(\ref{fluid:f.sol1})-\eqref{fluid:f.sol3} determine 
the metric function $f=f(r)$ and
\be
\label{explicitgammaprime}
\gamma'
=
\frac{2}{r\,f(r)}
\left[\gn\,m_{\rm B}'(r)
+\sqrt{a_0\,\gn\,m_{\rm B}(r)}
\right]
\ .
\ee
We examine for simplicity the case of baryonic matter localised 
inside a sphere of radius $R_{\rm B}\ll r_0$, so that the baryonic 
mass has a constant profile $m_{\rm B}(r)= m_{\rm B}$, for $r>R_{\rm B}$.
This approximation is good when we consider a galaxy at  distances 
much bigger than its bulk. Since we are now interested in scales 
$r\sim r_0\ll L$, we again neglect the DE terms, and the metric functions can 
be easily obtained from Eqs.~(\ref{explicitgammaprime})
and \eqref{fluid:f.sol1}--\eqref{fluid:f.sol3},
\be
\begin{array}{l}
f(r)
=
\strut\displaystyle{1-\frac{2\,\gn\,m_{\rm B}}{r}}
\\
\\
\gamma_{\rm DF}
=
\strut\displaystyle{
2\,K
\left[
\ln\left(\frac{r}{r_0}\right)
+\ln\!\left(1-\frac{2\,\gn\,m_{\rm B}}{r}\right)
\right]}
\ ,
\end{array}
\label{constmassgamma}
\ee
where $K=\sqrt{a_0\,\gn\,m_{\rm B}}$ and the integration constant 
was set in terms of the infrared scale $r_0$, which now represents the 
typical radius at which the ``dark force'' effects take place. 
\par
The non-vanishing function $\gamma_{\rm DF}$ represents the metric 
effects in our fluid description of the dark force.
Since our effective fluid description holds only for $r_0\lesssim r\ll L$,
we  neglect $\gamma_{\rm DF}$ for $r\lesssim r_0$ and $r\sim L$.
Most of the physical information about the rotation curves of the galaxies is 
contained in the weak-field approximation of the metric component 
$g_{00}=-f\,e^\gamma$.
At galactic scales, this corresponds to the regime $\gn\,m_{\rm B}\ll r\sim r_0\ll L$,
which also implies $\gamma_{\rm DF}\sim 0$.
Keeping only terms up to $\log^2(r/r_0)$ and $1/r^2$, we have
\be
\label{s1}
-g_{00}
&\simeq&
1-
\left(1+2\,K\right)\frac{2\,\gn\,m_{\rm B}}{r}
+ 2\,K\,\ln\!\left(\frac{r}{r_0}\right)
-
K\left(1+2\,K\right)\frac{4\,\gn\,m_{\rm B}}{r}\ln\!\left(\frac{r}{r_0}\right)
\ ,
\ee
where we exactly find the logarithmic corrections to the gravitational 
potential one expects at galactic scales, as MOND (or the Tully-Fisher 
relation) suggests~\cite{Milgrom:1983ca, McGaugh:2000sr, Battaner:2000ef}. 
Moreover, it contains the subleading $(1/r)\log (r/r_0)$ corrections, 
which have also been observed in galactic rotation
curves~\cite{Navarro:1995iw, Bertone:1900zza}.
A third feature of the above metric element is the presence of a small 
correction to the Newtonian potential, which can be seen 
as a modification of $\gn\,m_{\rm B}$, and depends on $a_0$ in $K$.
This correction is therefore of Machian character, but is tiny 
because $K$ is of order $10^{-6}$ for a spiral galaxy with 
$m_{\rm B}\sim 10^{11}\, m_{\odot}$, and of order $10^{-9}$ for a dwarf 
galaxy with $m_{\rm B}\sim 10^{7}\, m_{\odot}$. This effect is hence 
not detectable presently, owing to the uncertainties in the 
determination of the baryonic mass of the galaxies.
\par
Because of the competition between $\log(r/r_0)$ and $1/r$ terms (and 
also the dS term $r^2/L^2$ if one goes to distances comparable with 
the cosmological horizon) in the weak-field expansion, it is useful 
to introduce, beside $r_0$, the scales $r_1$ and $r_2$ representing 
the distances at which the MOND acceleration term equals respectively 
the Newtonian and the dS~term. Hence, our effective fluid description 
holds for $r_0<r<r_2$. The IR scale $r_0$ is the typical distance at 
which the rotation curves of galaxies deviate from the Newtonian 
prediction, $r_0\sim\sqrt{\gn\,m_{\rm B}\, L}$. In Verlinde's model 
of Ref.~\cite{Verlinde:2016toy}, the IR scale $r_0$ is determined by 
the competition between area and volume terms in the entropy, and is 
given by $r_0=\sqrt{2\,\gn\,m_{\rm B}\, L}$. In our case, we have 
$r_1= \sqrt{3}\, r_0$ and  $r_2=\sqrt{r_0\, L/(2\sqrt{3})}$. Notice 
that, as expected, $r_1\sim r_0$. The window 
in which the Newtonian contribution to the potential is not obscured by 
the logarithmic term is therefore very narrow. As specific examples, 
let us take the typical spiral and dwarf galaxies discussed above.
For the spiral galaxy, we have $r_0\simeq 6\,$Kpc, $r_1\simeq 10\,$Kpc, 
$r_2\simeq 10^3\,$Kpc. For the dwarf galaxy we have instead 
$r_0\simeq 80\,$pc, $r_1\simeq 130\,$pc, $r_2\simeq 300\,$pc.
\par
We have considered here only the case of a constant profile for the 
baryonic mass function outside a sphere of radius $R\ll r_0$.
However, Eqs.~(\ref{explicitgammaprime}) and \eqref{fluid:f.sol1}-\eqref{fluid:f.sol3}  
in principle allow for the determination of the metric for every 
given distribution of baryonic matter $m_{\rm B}=m_{\rm B}(r)$.
For instance, one can consider Jaffe's profile~\cite{Jaffe:1983iv} 
for the baryonic energy density $\varepsilon_{\rm B}={\tilde{A}}/{r^4}$, 
which corresponds to  $m_{\rm B}(r)=m_0-{A}/{r}$. We have checked that 
this profile reproduces the results for the case of a constant baryonic 
mass at large distances, as expected.
A detailed discussion of Jaffe's model will be presented in a forthcoming
paper.
\section{Conclusions and outlook}
\label{s:conc}
In this letter, we have proposed an effective fluid description in a 
GR framework for an infrared-modified theory of gravity. 
Using quite general assumptions and a microscopic description of 
the fluid in terms of a Bose-Einstein condensate of gravitons, we have 
found the static, spherically symmetric solution for the metric in 
terms of the Misner-Sharp mass function of baryonic matter and the 
fluid pressure.
In particular, we have shown that the additional component of the
acceleration at galactic scales can be completely attributed to the
radial pressure of the fluid, whose interpretation in the corpuscular model
is that this is part of the reaction of the condensate of gravitons to the
presence of baryonic matter.
Moreover, we have shown that our model correctly reproduces the leading
MOND $\log (r)$ and subleading $(1/r)\, \log( r)$ terms at galactic scales in 
the weak-field expansion of the potential.
Our model also predicts a tiny modification of the Newtonian potential at
galactic scales which is controlled by the cosmological acceleration.
\par
The next step in our analysis should be to test the model with 
observational data. Of particular interest are the situations 
where the predictions of our model are expected to differ from those 
of MOND and/or $\Lambda$CDM. For what concerns the dynamics of galactic 
systems, our model is testable for an isolated, spherically 
symmetric system. The most promising candidates are therefore spherical 
galaxies or isolated spherical dwarf galaxies and dwarf spheroidal satellite 
galaxies. On the other hand, as we have already seen, 
the point mass case leads to the same results of MOND. To have a 
first nontrivial test, {\em i.e.}~to look for significant differences 
between our model, MOND and $\Lambda$CDM we need 
to consider finite-size galaxies with a specific baryonic mass 
profile $m_B(r)$.
\par
At the present stage of development, the dynamics of galaxy clusters and
of systems that exhibit peculiar features as the external field effect of
MOND~\cite{Milgrom:2012xw} (like the Crater II dwarf satellite galaxy~\cite{McGaugh:2016jwx})
does not seem a suitable arena for testing the model.
In order to do that, extensions for composite systems and beyond the 
spherical symmetric approximation appear necessary.
\par
A different, but equally important challenge is represented by the study 
of the weak lensing effect at the level of galaxies and galaxy clusters. 
The well-defined form of the spacetime metric in~\eqref{explicitgammaprime} 
allows us to make predictions and, eventually, a direct comparison 
with results from both MOND and the $\Lambda$CDM model about the weak gravitational 
lensing measurements in galactic systems with static, spherically symmetric 
and isolated mass distributions. 
In order to do so, we need to choose our gravitational lenses 
to satisfy these criteria and to know the baryonic mass profile 
$m_B(r)$ of the system.
\section*{Acknowledgments}
This research was partially supported by INFN, research initiatives 
FLAG (R.C.~and A.G.), QUAGRAP (M.C.~and M.T.) and STEFI (W.M.). 
The work of R.C.~and A.G.~has been carried out in the framework 
of GNFM and INdAM and the COST action Cantata. 
\par
%
%
\providecommand{\href}[2]{#2}\begingroup\raggedright\endgroup

\end{document}